\documentclass[letterpaper,twocolumn,american,showpacs,pre,aps,superscriptaddress]{revtex4}
\usepackage[latin1]{inputenc}
\usepackage{bm}
\usepackage{multirow,amssymb,amsbsy,amsmath}
\usepackage{stmaryrd}
\usepackage{graphicx}
\usepackage{hyperref}
\makeatletter
\usepackage{pifont}
\makeatother

\begin{document}

\title{Stochastic analysis and simulation of spin star systems}

\author{Heinz-Peter Breuer}
\email{breuer@physik.uni-freiburg.de} \affiliation{Physikalisches
Institut, Universit\"at Freiburg, Hermann-Herder-Strasse 3,
D-79104 Freiburg, Germany}

\author{Francesco Petruccione}
\email{petruccione@ukzn.ac.za} \affiliation{School of Physics -
Quantum Research Group, University of KwaZulu-Natal, Westville
Campus, Private Bag X54001, Durban 4000, South Africa}

\date{\today}

\begin{abstract}
We discuss two methods of an exact stochastic representation of
the non-Markovian quantum dynamics of open systems. The first
method employs a pair of stochastic product vectors in the total
system's state space, while the second method uses a pair of state
vectors in the open system's state space and a random operator
acting on the state space of the environment. Both techniques lead
to an exact solution of the von Neumann equation for the density
matrix of the total system. Employing a spin star model describing
a central spin coupled to bath of surrounding spins, we perform
Monte Carlo simulations for both variants of the stochastic
dynamics. In addition, we derive analytical expression for the
expectation values of the stochastic dynamics to obtain the exact
solution for the density matrix of the central spin.
\end{abstract}


\maketitle

\section{Introduction}
The Markovian dynamics of an open quantum system $S$ which is
coupled to an environment $E$ \cite{TheWork} is conventionally
described by a master equation for the open system's density
matrix $\rho_S(t)$ with a generator in Lindblad form
\cite{GORINI,LINDBLAD}. It is a well-known feature
\cite{DALIBARD,DUM,GISIN,CARMICHAEL,PLENIO} of this type of master
equations that it yields a stochastic representation for
$\rho_S(t)$ in the form of an expectation value over an ensemble
of pure state vectors. This means that $\rho_S(t)$ can be
expressed by
\begin{equation} \label{EXPEC-STANDARD}
 \rho_S(t)={\mathrm{E}}(|\psi(t)\rangle\langle\psi(t)|),
\end{equation}
where $|\psi(t)\rangle$ is a stochastic state vector in the
Hilbert space of the open system and ${\mathrm{E}}$ denotes the
expectation value. The great advantage of the stochastic
representation consists in the fact that it leads to efficient
Monte Carlo techniques in which one propagates an ensemble of pure
state vectors in the open system's Hilbert space and estimates the
reduced density matrix through an appropriate ensemble average.

Recently, an exact stochastic treatment of non-Markovian quantum
dynamics has been proposed \cite{PDP-PRA,PDP-EPJD}. This method is
based on a representation of the density matrix $\rho(t)$ of the
total system through an expectation value of the form
\begin{equation} \label{EXPEC-NEW-1}
 \rho(t)={\mathrm{E}}(|\Phi_1(t)\rangle\langle\Phi_2(t)|).
\end{equation}
By contrast to the conventional approach one uses in this method a
pair of random product vectors
$|\Phi_1(t)\rangle=\psi_1(t)\otimes\chi_1(t)$ and
$|\Phi_2(t)\rangle=\psi_2(t)\otimes\chi_2(t)$ of the total system.
These product vectors follow independent stochastic time-evolution
equations that can be constructed in such a way that the average
over the probabilistic dynamics reproduces the exact Schr\"odinger
or von Neumann dynamics of the total system. As demonstrated in
\cite{PDP-PRA,PDP-EPJD} the evolution equations for the product
vectors $|\Phi_1(t)\rangle$ and $|\Phi_2(t)\rangle$ can be chosen
to be relatively simple time-local stochastic differential
equations, describing a piecewise deterministic process or a
diffusion process (Brownian motion) in Hilbert space. This means
that it is possible to design a representation of non-Markovian
quantum dynamics involving strong memory effects through a
Markovian unravelling by means of a pair of independent stochastic
product vectors.

This method bears several advantages. The stochastic differential
equations for the product vectors describe a Markovian random
process for which efficient numerical simulation algorithms are
known (see, e.~g., Ref.~\cite{TheWork} and references therein).
Since the method is based on a direct stochastic representation of
the full Schr\"odinger dynamics of the total system it does not
rely on the construction of an approximate effective master
equation for the reduced density matrix; it does not even require
the existence of such an equation. Furthermore, the method allows,
at least in principle, the treatment of arbitrary correlations in
the initial state. This follows from the fact that any initial
state $\rho(0)$ can be represented in the form
$\rho(0)={\mathrm{E}}(|\Phi_1(0)\rangle\langle\Phi_2(0)|)$, where
$|\Phi_1(0)\rangle$ and $|\Phi_2(0)\rangle$ are random product
states \cite{PDP-EPJD}. In particular, the method does not
presuppose that system and environment are initially in an
uncorrelated tensor product state. Finally, the technique not only
allows the determination of the reduced density matrix $\rho_S(t)$
but also of multitime quantum correlation functions of the open
system.

There is an important limitation in the applicability of the Monte
Carlo algorithms based on the stochastic representation
(\ref{EXPEC-NEW-1}) which is due to the behavior of the
statistical fluctuations. As shown in Ref.~\cite{PDP-EPJD} the
fluctuations of the process and, hence, also the statistical
errors of the Monte Carlo simulation may eventually grow
exponentially with time. Therefore, the method can generally be
expected to be feasible only for short time scales. However, it
must be noted that the stochastic dynamics of the product vectors
$|\Phi_1(t)\rangle$ and $|\Phi_2(t)\rangle$ is by no means unique,
i.~e., there exists an infinite number of stochastic evolution
equations for which the expectation value (\ref{EXPEC-NEW-1})
exactly represents the full system dynamics. Recently, this
freedom in the choice of an appropriate stochastic dynamics has
been employed to develop optimized Monte Carlo algorithms which
lead to a drastic reduction of the size of statistical errors
\cite{LACROIX}.

The structure of Eq.~(\ref{EXPEC-NEW-1}) is not the only
possibility of obtaining an exact stochastic representation for
the total density matrix. In fact, one can construct many other
random functionals whose expectation values lead to the desired
equation of motion. Here, we examine an alternative stochastic
formulation which employs a pair of random state vectors
$|\psi_1(t)\rangle$ and $|\psi_2(t)\rangle$ in the open system's
Hilbert space and a random operator $R_E(t)$ on the state space of
the environment:
\begin{equation} \label{EXPEC-NEW-2}
 \rho(t)={\mathrm{E}}(|\psi_1(t)\rangle\langle\psi_2(t)|\otimes
 R_E(t)).
\end{equation}
Again, one can construct an appropriate stochastic dynamics for
the state vectors $|\psi_1(t)\rangle$ and $|\psi_2(t)\rangle$ and
for the environmental operator $R_E(t)$ that guarantees that the
expectation value (\ref{EXPEC-NEW-2}) exactly satisfies the von
Neumann equation of the total system.

We start our considerations in Sec.~\ref{STOCH-REP} with a
description of the general concepts underlying the stochastic
representations given by Eqs.~(\ref{EXPEC-NEW-1}) and
(\ref{EXPEC-NEW-2}). The corresponding Monte Carlo simulation
techniques will be illustrated in Sec.~\ref{APPLICATIONS} with the
help of a spin star model, i.~e., a model of a central spin that
is coupled to a bath of surrounding spins \cite{BBP}. In addition
to performing numerical simulations, we derive analytical
expressions for the expectation values (\ref{EXPEC-NEW-1}) and
(\ref{EXPEC-NEW-2}) and relate these directly to the solution of
the Schr\"odinger equation for the total system. It turns out that
the method described by Eq.~(\ref{EXPEC-NEW-2}) is particularly
useful for the simulation of the dynamics for an infinite number
of bath spins. Some conclusions are drawn in Sec.~\ref{CONCLU}.

\section{Stochastic representations of non-Markovian quantum
         dynamics}\label{STOCH-REP}

\subsection{General theory}
We consider an open quantum system with Hilbert space
${\mathcal{H}}_S$ coupled to an environment with Hilbert space
${\mathcal{H}}_E$. The state space of the composite quantum system
is given by the tensor product space
${\mathcal{H}}_S\otimes{\mathcal{H}}_E$. Employing the interaction
picture we write the Hamiltonian describing the system-environment
coupling as follows,
\begin{equation} \label{HINT}
 H_I(t) = \sum_{\alpha} A_{\alpha}(t) \otimes B_{\alpha}(t).
\end{equation}
The $A_{\alpha}(t)$ and the $B_{\alpha}(t)$ are interaction
picture operators acting in ${\mathcal{H}}_S$ and
${\mathcal{H}}_E$, respectively. The corresponding von Neumann
equation for the density matrix $\rho(t)$ of the composite quantum
system is given by
\begin{equation} \label{NEUMANN}
 \frac{d}{dt} \rho(t) = -i [H_I(t),\rho(t)],
\end{equation}
where we set $\hbar = 1$.

Our aim is to construct a stochastic representation of the total
density matrix $\rho(t)$ in terms of the expectation value
\begin{equation} \label{EXPEC-GEN}
 \rho(t) = {\mathrm{E}}(R(t)).
\end{equation}
Here, $R(t)$ represents a random operator on the state space
${\mathcal{H}}_S\otimes{\mathcal{H}}_E$ of the total system. The
stochastic process governing the dynamics of this operator must be
constructed in such a way that the expectation value
(\ref{EXPEC-GEN}) satisfies the von Neumann equation
(\ref{NEUMANN}). It turns out that there are many possibilities of
constructing a stochastic representation which meets this
requirement. Of course, we do not seek just any stochastic
formulation, but the intention is to find a stochastic process
which leads to a considerable simplification of the representation
of the reduced density matrix
\begin{equation} \label{RHO-S}
 \rho_S(t) = {\mathrm{tr}}_E \rho(t)
 = {\mathrm{E}} \left( {\mathrm{tr}}_E R(t) \right)
\end{equation}
of the open system and which allows an efficient numerical
implementation of its time evolution (${\mathrm{tr}}_E$ denotes
the partial trace over ${\mathcal{H}}_E$). In the following we
discuss two (of many other) such possibilities, in which $R(t)$
follows a piecewise deterministic process (PDP) \cite{TheWork}.

\subsection{Stochastic process of the form
            $R=|\Phi_1\rangle\langle\Phi_2|$}\label{PDP1}
The first possibility of a stochastic representation in terms of a
PDP is given by taking $R(t)$ to be of the form
\begin{equation} \label{DEFR}
 R(t) = |\Phi_1(t)\rangle\langle\Phi_2(t)|,
\end{equation}
such that we have:
\begin{equation} \label{DEFPHINU}
 \rho(t) = {\mathrm{E}}(|\Phi_1(t)\rangle\langle\Phi_2(t)|).
\end{equation}
$|\Phi_1(t)\rangle$ and $|\Phi_2(t)\rangle$ represent a pair of
stochastic state vectors of the composite quantum system which are
chosen as direct products of system states $\psi_{\nu}(t)$ and
environmental states $\chi_{\nu}(t)$:
\begin{equation} \label{DIRECT-PRODUCT}
 |\Phi_\nu(t)\rangle = \psi_{\nu}(t) \otimes \chi_{\nu}(t),
 \qquad \nu = 1,2.
\end{equation}
In view of Eqs.~(\ref{DEFPHINU}) and (\ref{DIRECT-PRODUCT}) the
reduced density matrix $\rho_S(t)$ [see Eq.~(\ref{RHO-S})] can be
expressed in terms of the expectation value:
\begin{equation} \label{RHOS}
 \rho_S(t) = {\mathrm{E}} \left(
 |\psi_1(t)\rangle\langle\psi_2(t)|
 \langle \chi_2(t) | \chi_1(t) \rangle \right).
\end{equation}
By contrast to the standard stochastic unravelling of the dynamics
of open quantum systems, this representation employs an average
over the product of two quantities, namely the dyadic
$|\psi_1\rangle\langle\psi_2|$ formed by a pair $\psi_1$, $\psi_2$
of state vectors of the open system, and the scalar product
$\langle \chi_2 | \chi_1 \rangle$ of a corresponding pair of
environment states.

According to the ansatz (\ref{DIRECT-PRODUCT}) the states
$|\Phi_\nu(t)\rangle$ of the total system are direct products at
any time $t$, which greatly simplifies the representation of the
states of the system and the simulation of its dynamics. Of
course, the exact states are generally entangled. This shows that
the dynamics of the $|\Phi_\nu(t)\rangle$ cannot be described by a
deterministic time evolution. However, as demonstrated in
\cite{PDP-EPJD,PDP-PRA} it is possible to reproduce the dynamics
of the total density matrix with the help of a random Markov
process. An appropriate system of stochastic differential
equations for the $\psi_{\nu}$ and the $\chi_{\nu}$ is given by
\begin{equation}
 d\psi_{\nu}(t) = \sum_{\alpha}
 \left( \frac{-i||\psi_{\nu}||}{||A_{\alpha}(t)\psi_{\nu}||}A_{\alpha}(t)-I\right)
 \psi_{\nu} dN_{\alpha\nu}(t) \label{STOCH1}
\end{equation}
and
\begin{eqnarray}
 d\chi_{\nu}(t) &=& \Gamma_{\nu}(t) \chi_{\nu} dt \\
 &+&  \sum_{\alpha}
 \left( \frac{||\chi_{\nu}||}{||B_{\alpha}(t)\chi_{\nu}||}B_{\alpha}(t)-I\right)
 \chi_{\nu} dN_{\alpha\nu}(t). \nonumber \label{STOCH2}
\end{eqnarray}
where $I$ denotes the unit operator. The quantities
$dN_{\alpha\nu}$ are random Poisson increments which satisfy
\begin{equation} \label{EDN}
 {\mathrm{E}}(dN_{\alpha\nu}(t)) = \Gamma_{\alpha\nu}(t)dt
\end{equation}
and
\begin{equation} \label{DNDN}
 dN_{\alpha\nu}(t) dN_{\beta\mu}(t) =
 \delta_{\alpha\beta} \delta_{\nu\mu} dN_{\alpha\nu}(t).
\end{equation}
The corresponding rates are given by
\begin{equation} \label{DEFGANU}
 \Gamma_{\alpha\nu}(t) =
 \frac{||A_{\alpha}(t)\psi_{\nu}||\cdot||B_{\alpha}(t)\chi_{\nu}||}
 {||\psi_{\nu}||\cdot||\chi_{\nu}||},
\end{equation}
and we have defined the total rates
\begin{equation} \label{DEFGNU}
 \Gamma_{\nu}(t) \equiv \sum_{\alpha} \Gamma_{\alpha\nu}(t).
\end{equation}

In view of Eq.~(\ref{DNDN}) the stochastic increments
$dN_{\alpha\nu}(t)$ take on the possible values $0$ or $1$.
According to Eq.~(\ref{EDN}) the case $dN_{\alpha\nu}(t)=1$ occurs
with probability $\Gamma_{\alpha\nu}dt$. Under the condition that
$dN_{\alpha\nu}(t)=1$ for a particular $\alpha$ and $\nu$, the
other increments vanish, and Eqs.~(\ref{STOCH1}) and
(\ref{STOCH2}) imply that for this particular $\alpha$ and $\nu$
the state vectors $\psi_{\nu}$ and $\chi_{\nu}$ perform the
instantaneous jumps
\begin{equation} \label{QJUMPS}
 \psi_{\nu} \rightarrow
 \frac{-i||\psi_{\nu}||}{||A_{\alpha}\psi_{\nu}||} A_{\alpha}\psi_{\nu},
 \qquad
 \chi_{\nu} \rightarrow
 \frac{||\chi_{\nu}||}{||B_{\alpha}\chi_{\nu}||} B_{\alpha}\chi_{\nu}.
\end{equation}
Note that these jumps preserve the norm of the state vectors.
Under the condition that all Poisson increments vanish, that is
$dN_{\alpha\nu}(t)=0$ for all $\alpha$ and $\nu$, we have
$d\psi_{\nu}(t)=0$ and $d\chi_{\nu}(t)=\Gamma_{\nu}\chi_{\nu}dt$.
This means that $\psi_{\nu}$ remains unchanged during $dt$, while
$\chi_{\nu}$ follows a linear drift.

Summarizing, $\psi_{\nu}(t)$ is a pure, norm-conserving jump
process, whereas $\chi_{\nu}(t)$ is a PDP with norm-conserving
jumps and a linear drift. It is demonstrated in \cite{PDP-EPJD}
that any initial density matrix of the total system can be
represented in the form
$\rho(0)={\mathrm{E}}(|\Phi_1(0)\rangle\langle\Phi_2(0)|)$, and
that the expectation value (\ref{DEFPHINU}) exactly satisfies the
von Neumann equation (\ref{NEUMANN}). These facts enable us to
simulate the full non-Markovian quantum dynamics through a Monte
Carlo simulation of the stochastic differential equations
(\ref{STOCH1}) and (\ref{STOCH2}).

\subsection{Stochastic process of the form
            $R=|\psi_1\rangle\langle\psi_2|\otimes R_E$}\label{PDP2}
The second possibility of a stochastic representation is obtained
if we take $R(t)$ to be of the form
\begin{equation} \label{DEFR2}
 R(t) = |\psi_1(t)\rangle\langle\psi_2(t)|\otimes R_E(t),
\end{equation}
such that we have:
\begin{equation} \label{DEFRHO2}
 \rho(t) = {\mathrm{E}}(|\psi_1(t)\rangle\langle\psi_2(t)|\otimes R_E(t)).
\end{equation}
In this case we represent $R(t)$ through a pair $\psi_1$, $\psi_2$
of state vectors of the open system and a random operator $R_E$ on
the state space of the environment. The reduced density matrix of
the open system can thus be written as
\begin{equation} \label{RHOS2}
 \rho_S(t) = {\mathrm{E}} \left(
 |\psi_1(t)\rangle\langle\psi_2(t)| {\textrm{tr}}_E R_E(t) \right).
\end{equation}
An appropriate system of stochastic differential equation which
reproduces the exact von Neumann dynamics for the expectation
value (\ref{DEFRHO2}) is given by
\begin{eqnarray}
 d\psi_{\nu}(t) &=& \sum_{\alpha}
 \left( -iL_{\alpha\nu}A_{\alpha}-I\right)
 \psi_{\nu}(t) dN_{\alpha\nu}(t), \label{STOCH1-RE} \\
 dR_E(t) &=& \Gamma R_E(t) dt \nonumber \\
 &~& + \sum_{\alpha}
 \left( M_{\alpha 1} B_{\alpha}-I\right) R_E(t) dN_{\alpha 1}(t)
 \nonumber \\
 &~& + \sum_{\alpha} R_E(t)
 \left( M_{\alpha 2} B^{\dagger}_{\alpha}-I\right) dN_{\alpha 2}(t).
 \label{STOCH2-RE}
\end{eqnarray}
The Poisson increments $dN_{\alpha\nu}$ satisfy Eqs.~(\ref{EDN})
and (\ref{DNDN}), and the transition rates are given by
\begin{equation} \label{DEFGANU-RE}
 \Gamma_{\alpha\nu} = \frac{1}{L_{\alpha\nu}M_{\alpha\nu}},
\end{equation}
and
\begin{eqnarray}
 \Gamma_{\nu} &=& \sum_{\alpha}\Gamma_{\alpha\nu}, \\
 \Gamma &=& \Gamma_1 + \Gamma_2. \label{DEFGNU-RE}
\end{eqnarray}
The quantities $L_{\alpha\nu}$ and $M_{\alpha\nu}$ are real and
positive functionals of $\psi_1$, $\psi_2$ and $R_E$. One has a
great freedom in the choice of these functionals, the only
restriction being the positivity. A definite choice will be made
in the example below.

We observe that again the $\psi_{\nu}$ follow a pure jump process.
If $dN_{\alpha\nu}(t)=1$ for a particular pair of indices $\alpha$
and $\nu$, which happens with probability $\Gamma_{\alpha\nu}dt$,
the state vector $\psi_{\nu}$ undergoes the jump
\begin{equation} \label{QJUMPS-R-psi}
 \psi_{\nu} \rightarrow
 -iL_{\alpha\nu}A_{\alpha}\psi_{\nu}.
\end{equation}
At the same time $R_E$ carries out the jump
\begin{equation}
 R_E \rightarrow M_{\alpha 1}B_{\alpha} R_E
\end{equation}
if $\nu=1$ and the jump
\begin{equation}
 R_E \rightarrow R_E M_{\alpha 2}B^{\dagger}_{\alpha}
\end{equation}
if $\nu=2$. Thus, for $\nu=1$ the operator $B_{\alpha}$ acts from
the left on $R_E$, while for $\nu=2$ the adjoint operator
$B^{\dagger}_{\alpha}$ acts from the right on $R_E$. Under the
condition that all Poisson increments vanish, which occurs with
probability
\begin{equation}
 1-\sum_{\alpha\nu}\Gamma_{\alpha\nu}dt = 1-\Gamma dt,
\end{equation}
we have $d\psi_{\nu}(t)=0$ and $dR_E(t)=\Gamma R_E(t)dt$, i.e.,
the $\psi_{\nu}$ are left unchanged during $dt$ while the
environment matrix $R_E(t)$ follows a linear drift.

Like in the case of the process constructed in Sec.~\ref{PDP1} it
is easy to design an appropriate Monte Carlo algorithm for the
stochastic differential equations (\ref{STOCH1-RE}) and
(\ref{STOCH2-RE}). We note that in both cases the random matrix
$R(t)$ has the structure of a tensor product, which considerably
reduces the complexity of the problem. As a consequence of
Eqs.~(\ref{RHOS}) and (\ref{RHOS2}) the environmental states enter
the expectation value for the reduced density matrix $\rho_S(t)$
only through the scalar product
$\langle\chi_2(t)|\chi_1(t)\rangle$ or through the trace
${\textrm{tr}}_E R_E(t)$. To simulate the non-Markovian dynamics
of an open system with these algorithms it thus suffices to record
during the simulation of the process the various jumps and their
moments of occurrence. At any time $t$ the scalar product
$\langle\chi_2(t)|\chi_1(t)\rangle$ or the trace ${\textrm{tr}}_E
R_E(t)$ can then be expressed in terms of certain correlation
functions of the environmental operators $B_{\alpha}$. For many
system-environment models the latter are known explicitly. This
fact greatly facilitates the numerical implementation of the
stochastic method. An example is discussed in Sec.~\ref{EXAMPLE-2}
(see, in particular, Eq.~(\ref{TRACE-RE})).

\section{Applications}\label{APPLICATIONS}

\subsection{The spin star model}\label{SPINSTAR}
As a simple but instructive example of the Monte Carlo method, we
investigate in the following a spin star model described by the
time independent interaction Hamiltonian
\begin{equation} \label{hamiltonian}
 H = \frac{2A}{\sqrt{N}} \left( \sigma_{+}J_{-}+\sigma
 _{-}J_{+}\right).
\end{equation}
The Pauli spin operator of the central spin, constituting the open
system, is denoted by $\vec{\sigma}$ with corresponding raising
and lowering operators $\sigma_{\pm}=\frac{1}{2}(\sigma_1 \pm
i\sigma_2)$. The central spin couples to $N$ environment spins
with Pauli spin operators $\vec{\sigma}^{(i)}$, $i=1,2,\ldots,N$,
through the raising and lowering operators
\begin{equation}
 J_{\pm } \equiv \sum^{N}_{i=1} \sigma ^{(i)}_{\pm}
 = \frac{1}{2} \sum_{i=1}^{N}
 \left( \sigma^{(i)}_{1} \pm i\sigma^{(i)}_{2}\right)
\end{equation}
of the total angular momentum $\vec{J}$ of the environment. The
initial state of the total system at time $t=0$ is taken to be a
product state $\rho(0)=\rho_{S}(0)\otimes\rho_E(0)$, where the
reduced density matrix $\rho_S(0)$ of the central spin may be an
arbitrary, possibly mixed state. The spin bath is assumed to be in
an unpolarized infinite temperature initial state
\begin{equation} \label{INIT-MIX}
 \rho_E(0) = 2^{-N} I_E,
\end{equation}
where $I_E$ denotes the unit matrix in ${\mathcal{H}}_E$.

This model can be solved analytically \cite{BBP}. We express the
solution for the reduced density matrix in terms of the components
of the Bloch vector
\begin{equation}
 \vec{v}(t) = {\textrm{tr}}\left\{ \sigma \rho(t) \right\}
\end{equation}
which are related to the reduced density matrix by
\begin{eqnarray}
 \rho _{S}(t) &=& \left( \begin{array}{cc}
 \rho_{++}(t) & \rho_{+-}(t)\\
 \rho_{-+}(t) & \rho_{--}(t)
 \end{array}\right) \nonumber \\
 &=&  \frac{1}{2}\left( \begin{array}{cc}
 1+v_{3}(t) & v_{1}(t)-iv_{2}(t)\\
 v_{1}(t)+iv_{2}(t) & 1-v_{3}(t)
 \end{array}\right). \label{relationship_bloch_dens}
\end{eqnarray}
The components $v_3(t)$ and $v_{\pm}(t)\equiv\frac{1}{2}(v_{1}\pm
iv_{2})$ are then given by the explicit expressions:
\begin{eqnarray}
 \!\!\!\!\!\!\!\!\! \frac{v_3(t)}{v_3(0)} &=& \sum_{j,m} P(j,m)
 \cos\left[2\Gamma(j,m)t\right], \label{SOL3} \\
 \!\!\!\!\!\!\!\!\! \frac{v_{\pm(t)}}{v_{\pm}(0)} &=&
 \sum_{j,m} P(j,m)
 \cos\left[\Gamma(j,m)t\right]\cos\left[\Gamma(j,-m)t\right],
 \label{SOL12}
\end{eqnarray}
where
\begin{equation} \label{PJM}
 P(j,m) = \frac{1}{2^N}
 \left[\binom{N}{\frac{N}{2}+j}-\binom{N}{\frac{N}{2}+j+1}\right],
\end{equation}
and
\begin{equation} \label{GJM}
 \Gamma(j,m) = 2A\sqrt{\frac{j(j+1)-m(m+1)}{N}}.
\end{equation}
These expressions may be obtained by solving the Schr\"odinger
equation of the model with the help of the fact that the manifolds
spanned by the states $|+\rangle \otimes |j,m\rangle$ and
$|-\rangle \otimes |j,m+1\rangle$ are invariant under the time
evolution. Here, $|\pm\rangle$ are the eigenstates of $\sigma_3$
with corresponding eigenvalues $\pm 1$, and $|j,m\rangle$ denotes
an eigenstate of the square $\vec{J}\,^2$ of the angular momentum
$\vec{J}$ of the spin bath and of its 3-component $J_3$ with
respective eigenvalues $j(j+1)$ and $m$.

The quantity $P(j,m)$ defined in Eq.~(\ref{PJM}) is the
probability of finding the quantum numbers $j$ and $m$ in the
initial mixture representing the state (\ref{INIT-MIX})
\cite{WESENBERG}. As usual, for $N$ even $j$ takes on the values
$j=0,1,2,\ldots,N/2$, and the values $j=1/2,3/2,\ldots,N/2$ if $N$
is odd. For a given $j$ the quantum number $m$ takes the values
$-j,-j+1,\ldots,+j$. It is easy to check that the probability
distribution $P(j,m)$ is normalized as follows,
\begin{equation}
 \sum_{j,m} P(j,m) = 1.
\end{equation}

In the limit $N\rightarrow\infty$ of an infinite number of bath
spins the above formulae lead to the asymptotic expressions
\cite{BBP}:
\begin{eqnarray}
 \frac{v_3(t)}{v_3(0)} &=&  1 + 2g(t), \label{F3infty}\\
 \frac{v_{\pm(t)}}{v_{\pm}(0)} &=& 1 + g(t), \label{F12infty}
\end{eqnarray}
where
\begin{equation} \label{Gtinfty}
 g(t) \equiv -\frac{\sqrt{\pi}}{2} x e^{-x^2} \textrm{erfi}(x)
 = \sum^{\infty}_{k=1}\frac{(-1)^k k!}{2(2k)!} (2x)^{2k},
\end{equation}
with $x\equiv\sqrt{2}At$. The function $\textrm{erfi}(x)$ denotes
the imaginary error function, which is a real-valued function
defined by
\begin{equation}
 \textrm{erfi}(x)\equiv \frac{\textrm{erf}(ix)}{i}
 = \frac{2}{\sqrt{\pi }}\sum ^{\infty}_{k=0}\frac{x^{2k+1}}{k!(2k+1)}.
\end{equation}

\subsection{The representation $R=|\Phi_1\rangle\langle\Phi_2|$}
\label{EXAMPLE-1}

We first illustrate the stochastic representation for the process
defined in Sec.~\ref{PDP1}. To this end, the initial state
(\ref{INIT-MIX}) is realized through a mixture of the states
$|j,m\rangle$, where the quantum numbers $j$ and $m$ follow the
joint probability distribution $P(j,m)$ given in Eq.~(\ref{PJM}).
To analyze the process we therefore have to describe the
stochastic evolution of the initial states
\begin{equation}
 |\Phi_1(0)\rangle = |\Phi_2(0)\rangle
 = |+\rangle \otimes |j,m\rangle, \label{INIT1}
\end{equation}
or
\begin{equation}
 |\Phi_1(0)\rangle = |\Phi_2(0)\rangle
 = |-\rangle \otimes |j,m\rangle, \label{INIT2}
\end{equation}
from which we can reconstruct the reduced density matrix of the
central spin.

Let us consider first the initial state (\ref{INIT1}). According
to the interaction Hamiltonian (\ref{hamiltonian}) the index
$\alpha$ in Eq.~(\ref{HINT}) assumes two values $\alpha=\pm$ with
corresponding time independent operators
\begin{equation}
 A_{\pm} = \sigma_{\pm}, \qquad B_{\pm} = \frac{2A}{\sqrt{N}}J_{\mp}.
\end{equation}
The jumps of the process thus take the form
\begin{equation} \label{QJUMPS-SPIN}
 \psi_{\nu} \rightarrow
 \frac{-i||\psi_{\nu}||}{||\sigma_{\pm}\psi_{\nu}||} \sigma_{\pm}\psi_{\nu},
 \qquad
 \chi_{\nu} \rightarrow
 \frac{||\chi_{\nu}||}{||J_{\mp}\chi_{\nu}||} J_{\mp}\chi_{\nu}.
\end{equation}
Since $\psi_{\nu}(0)=|+\rangle$ the states $\psi_{\nu}$ jump
between the states $|+\rangle$ and $|-\rangle$, whereby each jump
contributes an additional factor of $(-i)$. Let us denote the
number of jumps of $|\Phi_{\nu}\rangle$ during the time interval
from $0$ to $t$ by $n_{\nu}=n_{\nu}(t)$. We then have
$\psi_{\nu}(t)=(-i)^{n_{\nu}}|+\rangle$ if $n_{\nu}$ is even, and
$\psi_{\nu}(t)=(-i)^{n_{\nu}}|-\rangle$ if $n_{\nu}$ is odd. The
corresponding environment states are given by
\begin{equation} \label{CHINU1-SPIN}
 \chi_{\nu}(t) = |j,m\rangle \, e^{\Gamma(j,m)t}
\end{equation}
for even $n_{\nu}$, and by
\begin{equation} \label{CHINU2-SPIN}
 \chi_{\nu}(t) = |j,m+1\rangle \, e^{\Gamma(j,m)t}
\end{equation}
for odd $n_{\nu}$. The rates $\Gamma_{\pm}$ of the jumps are
determined as follows,
\begin{eqnarray} \label{RATE+}
 \Gamma_- &=&
 \frac{2A}{\sqrt{N}}||J_+|j,m\rangle || = \Gamma(j,m), \\
 \Gamma_+ &=&
 \frac{2A}{\sqrt{N}}||J_-|j,m+1\rangle || = \Gamma(j,m).
\end{eqnarray}
Thus we have $\Gamma_+=\Gamma_-=\Gamma(j,m)$, where $\Gamma(j,m)$
is given by Eq.~(\ref{GJM}). The deterministic drift of the
process therefore yields a factor $\exp[\Gamma(j,m)t]$, which has
already been taken into account in Eqs.~(\ref{CHINU1-SPIN}) and
(\ref{CHINU2-SPIN}).

On using this information we can determine the dynamics of the
populations of the reduced density matrix. Considering
$\rho_S(0)=|+\rangle\langle +|$, we have by virtue of
Eq.~(\ref{RHOS}):
\begin{eqnarray} \label{EXPEC1-SPIN}
 \rho_{++}(t) &=& \langle + | \rho_S(t) | + \rangle \\
 &=& {\mathrm{E}} \left(
 \langle + |\psi_1(t)\rangle \langle\psi_2(t)| + \rangle
 \langle \chi_2(t) | \chi_1(t) \rangle \right) \nonumber \\
 &=& {\mathrm{E}} \left( w(n_1,n_2) (-i)^{n_1} (+i)^{n_2}
 e^{2\Gamma(j,m)t} \right). \nonumber
\end{eqnarray}
Of course, a given realization of the process contributes to the
expectation value only if $n_1$ and $n_2$ are even. This fact is
accounted for by the first factor $w(n_1,n_2)$ which is defined to
be equal to $1$ if $n_1$ and $n_2$ are even, and equal to zero
otherwise. The second and the third factor under the expectation
value take into account the jumps of $\psi_1$ (factor
$(-i)^{n_1}$) and of $\psi_2$ (factor
$\left[(-i)^{n_2}\right]^{\ast}=(+i)^{n_2}$). Finally, the
exponential function represents the contributions from the scalar
product $\langle \chi_2(t)|\chi_1(t) \rangle$ which, according to
Eq.~(\ref{CHINU1-SPIN}), equals $\exp[2\Gamma(j,m)t]$.

It is clear from the general theory outlined in Sec.~\ref{PDP1}
that Eq.~(\ref{EXPEC1-SPIN}) is an exact representation of the
populations. Nevertheless, it might be instructive to see
explicitly how the exact solution (\ref{SOL3}) for the 3-component
of the Bloch vector emerges from the expectation value
(\ref{EXPEC1-SPIN}). To this end, we note that the states
$|\Phi_{\nu}\rangle$ evolve independently and that the transition
rates of the process are time independent. This implies that the
random numbers $n_1(t)$ and $n_2(t)$ follow independent Poisson
distributions with the same mean value of $\Gamma(j,m)t$:
\begin{equation} \label{PK1K2}
 P(n_{\nu},t) = \frac{[\Gamma(j,m)t]^{n_{\nu}}}{n_{\nu}!}
 e^{-\Gamma(j,m)t}.
\end{equation}
The expectation value (\ref{EXPEC1-SPIN}) therefore becomes
\begin{eqnarray} \label{EXPEC2-SPIN}
 \rho_{++}(t) &=& \sum_{j,m} P(j,m) \sum_{n_1,n_2} (-i)^{n_1}
 (+i)^{n_2} \nonumber \\
 &~& \qquad \times e^{2\Gamma(j,m)t} P(n_1,t) P(n_2,t).
\end{eqnarray}
The first sum extends over all possible values of the quantum
numbers $j$ and $m$ occurring in the initial state (see
Sec.~\ref{SPINSTAR}), while the second sum runs over all
$n_1,n_2=0,2,4,\ldots$. Substituting the expression (\ref{PK1K2})
into Eq.~(\ref{EXPEC2-SPIN}), we get
\begin{eqnarray} \label{EXPEC3-SPIN}
 \rho_{++}(t) &=& \sum_{j,m} P(j,m)
 \sum_{n_1,n_2} (-i)^{n_1} (+i)^{n_2}
 \nonumber \\
 &~& \qquad \times
 \frac{[\Gamma(j,m)t]^{n_1}}{n_1!}\frac{[\Gamma(j,m)t]^{n_2}}{n_2!}
 \nonumber \\
 &=& \sum_{j,m} P(j,m) \cos^2(\Gamma(j,m)t).
\end{eqnarray}
Using, finally, the relation $v_3(t)=2\rho_{++}-1$ we see that
Eq.~(\ref{EXPEC3-SPIN}) leads to the exact expression (\ref{SOL3})
for the 3-component of the Bloch vector. Thus we see explicitly
that the stochastic process indeed reproduces correctly the exact
time evolution of the system.

In a similar way one finds the coherence $v_-(t)=\rho_{+-}(t)$ of
the central spin. To this end, we have to consider also the
initial state (\ref{INIT2}). The resulting process is essentially
the same as above, with the only difference that now
$\Gamma_+=\Gamma_-=\Gamma(j,-m)$. To find the expectation value
representing $v_-(t)$ we have to use the initial states
$|\Phi_{1}(0)\rangle=|+\rangle \otimes |j,m\rangle$ and
$|\Phi_{2}(0)\rangle=|-\rangle \otimes |j,m\rangle$. With the
initial condition $v_-(0)=1$ we then have
\begin{eqnarray} \label{EXPEC1-SPIN-COH}
 v_-(t) &=& \langle + | \rho_S(t) | - \rangle \\
 &=& {\mathrm{E}} \left(
 \langle + |\psi_1(t)\rangle \langle\psi_2(t)| - \rangle
 \langle \chi_2(t) | \chi_1(t) \rangle \right) \nonumber \\
 &=& {\mathrm{E}} \left( w(n_1,n_2) (-i)^{n_1} (+i)^{n_2}
 e^{(\Gamma(j,m)+\Gamma(j,-m))t} \right). \nonumber
\end{eqnarray}
It is easy to verify that this expectation value leads to the
exact expression (\ref{SOL12}) for the coherence of the central
spin.

\begin{figure}[htb]
\includegraphics[width=\linewidth]{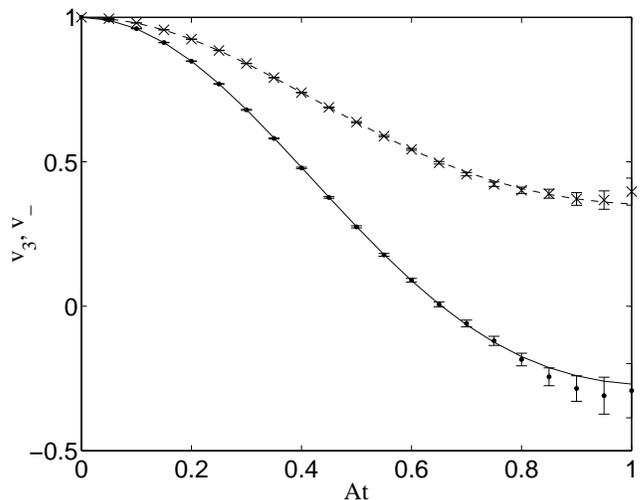}
\caption{\label{figure1} Monte Carlo simulation of
Eqs.~(\ref{STOCH1}) and (\ref{STOCH2}) using a sample of $10^7$
realizations: The 3-component $v_3=2\rho_{++}-1$ (dots and
errorbars) of the Bloch vector and the coherence $v_-=\rho_{+-}$
(crosses and errorbars) of the central spin coupled to a bath of
$10^2$ spins through the Hamiltonian (\ref{hamiltonian}). The
continuous line and the broken line show the analytical solution
given by Eqs.~(\ref{SOL3}) and (\ref{SOL12}), respectively.}
\end{figure}

Figure \ref{figure1} shows an example of a Monte Carlo simulation
of the stochastic process defined by the differential equations
(\ref{STOCH1}) and (\ref{STOCH2}). As can be seen, the Monte Carlo
simulation reproduces the exact solution with high accuracy over
the range of time shown. The figure also indicates the growth of
the size of the statistical errors which have been estimated from
the sample of realizations generated.

Beyond the point $At \approx 1$ the statistical errors strongly
increase. This is a typical feature of the Monte Carlo simulation
method which also appears in many other models. In fact, if one
measures the size of the statistical errors by means of the
Hilbert-Schmidt distance between the random operator $R(t)$ and
its mean value $\rho(t)$, one can show \cite{PDP-EPJD} that for
large times the fluctuations grow roughly as $\exp(2\Gamma_0t)$,
where $\Gamma_0$ represents an upper bound for the rates
$\Gamma_{\nu}$. As can be seen from the above example this
exponential increase of the errors is mainly due to the
corresponding increase of the norm of the environmental states
$\chi_{\nu}$.

\subsection{The representation $R=|\psi_1\rangle\langle\psi_2|\otimes R_E$}
\label{EXAMPLE-2}

Let us now analyze the process defined in Sec.~\ref{PDP2}, which
is particularly suited to simulate the limit of an infinite number
$N$ of bath spins. We choose the quantities $L_{\alpha\nu}$ as
follows:
\begin{equation}
 L_{\alpha\nu} = \frac{||\psi_{\nu}||}{||A_{\alpha}\psi_{\nu}||}.
\end{equation}
In our example we then have $L_{\alpha\nu}=1$. Thus, $\psi_{\nu}$
again jumps between states $|+\rangle$ and $|-\rangle$, whereby
each jump contributes a factor of $(-i)$. Hence, we have
\begin{equation} \label{EXPEC1-SPIN-2}
 \rho_{++}(t) = {\mathrm{E}} \left(
 w(n_1,n_2) (-i)^{n_1} (+i)^{n_2}
 {\textrm{tr}}_E R_E(t) \right).
\end{equation}

The aim is now to determine the trace over the random
environmental operator $R_E(t)$. This will be done in the limit
$N\rightarrow\infty$. In this limit we have for a fixed number
$k=0,1,2,\ldots$ \cite{BBP}
\begin{equation} \label{ASYMPT}
 \langle \left( J_+J_- \right)^k \rangle =
 k! \left( \frac{N}{2} \right)^k,
\end{equation}
where we define
\begin{equation} \label{DEF-ANGULAR}
 \langle {\mathcal{O}} \rangle \equiv
 {\mathrm{tr}}_E \left\{ {\mathcal{O}} \rho_E(0) \right\} =
 2^{-N} {\textrm{tr}}_E {\mathcal{O}}
\end{equation}
for any bath operator ${\mathcal{O}}$. We further choose the jump
rates $\Gamma_1=\Gamma_2=\sqrt{2}A$ (the factor $\sqrt{2}$ is
introduced for convenience), such that
$\Gamma=\Gamma_1+\Gamma_2=2\sqrt{2}A$. The drift contribution to
$R_E(t)$ is therefore given by $\exp(2\sqrt{2}At)$. The jumps of
the random matrix $R_E$ take the form
\begin{equation}
 R_E \rightarrow \frac{1}{\sqrt{2}A} B_{\pm} R_E
 = \sqrt{\frac{2}{N}} J_{\mp} R_E,
\end{equation}
or
\begin{equation}
 R_E \rightarrow \frac{1}{\sqrt{2}A} R_E B^{\dagger}_{\pm}
 = \sqrt{\frac{2}{N}} R_E J_{\pm}.
\end{equation}
Since $R_E(0)$ is proportional to the identity and since the order
of application of the operators $J_{\pm}$ is irrelevant in the
limit $N\rightarrow\infty$, we conclude that
\begin{equation} \label{TRACE-RE}
 {\textrm{tr}}_E R_E(t) = \left( \sqrt{\frac{2}{N}} \right)^{2k}
 \langle \left( J_+J_- \right)^k \rangle \, e^{\Gamma t},
\end{equation}
where we have defined $k=(n_1+n_2)/2$, assuming that both $n_1$
and $n_2$ are even. Employing Eq.~(\ref{ASYMPT}) we therefore get
\begin{equation}
 {\textrm{tr}}_E R_E(t) = k! e^{\Gamma t}.
\end{equation}
Hence, the expectation value (\ref{EXPEC1-SPIN-2}) becomes:
\begin{equation} \label{EXPEC2-SPIN-2}
 \rho_{++}(t) = {\mathrm{E}} \left(
 w(n_1,n_2) (-1)^k k! e^{\Gamma t} \right).
\end{equation}

\begin{figure}[htb]
\includegraphics[width=\linewidth]{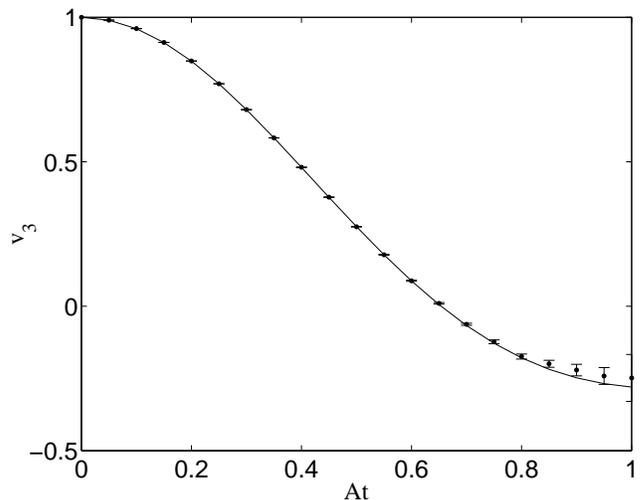}
\caption{\label{figure2} The 3-component $v_3=2\rho_{++}-1$ of the
Bloch vector of the central spin coupled to an infinite number of
bath spins through the Hamiltonian (\ref{hamiltonian}). Dots and
errorbars: Monte Carlo simulation based on the stochastic
differential equations (\ref{STOCH1-RE}) and (\ref{STOCH2-RE})
with $10^7$ realizations. Continuous line: Analytical solution
given by Eqs.~(\ref{F3infty}) and (\ref{Gtinfty}).}
\end{figure}

It is again instructive to see explicitly how this expression
leads to the formulae (\ref{F3infty}) and (\ref{F12infty}). Using
the fact that $n_1(t)$ and $n_2(t)$ are independent and follow
Poisson distributions with mean value $\sqrt{2}At$, one finds
\begin{eqnarray*}
 \rho_{++}(t) &=& \sum_{n_1,n_2}
 \frac{(\sqrt{2}At)^{n_1}}{n_1!}\frac{(\sqrt{2}At)^{n_2}}{n_2!}
 (-1)^k k! \\
 &=& \sum_{k=0}^{\infty} \frac{(-1)^k k!}{(2k)!}
 (\sqrt{2}At)^{2k} \sum_{n_1} \binom{2k}{n_1}.
\end{eqnarray*}
We recall that the sum in the first line extends over the values
$n_1,n_2=0,2,4,\ldots$, and that we use the definition
$k=(n_1+n_2)/2$. The second sum in the second line runs over the
values $n_1=0,2,\ldots,2k$. This sum is found to be equal to $1$
for $k=0$ and equal to $2^{2k-1}$ for $k=1,2,3,\ldots$. Thus we
obtain
\begin{equation}
 \rho_{++}(t) = 1 + \sum_{k=1}^{\infty} \frac{(-1)^k k!}{2(2k)!}
 (2\sqrt{2}At)^{2k} \equiv 1+g(t),
\end{equation}
from which we get $v_3(t)=2\rho_{++}(t)-1=1+2g(t)$, where the
function $g(t)$ has been introduced in Eq.~(\ref{Gtinfty}). A
similar reasoning leads to the relation $v_{\pm}(t)=1+g(t)$. These
results coincide with those obtained from the solution of the
Schr\"odinger equation [see Sec.~\ref{SPINSTAR}].

Figure \ref{figure2} shows the results of a Monte Carlo simulation
of the stochastic process defined by the differential equations
(\ref{STOCH1-RE}) and (\ref{STOCH2-RE}). The Monte Carlo
simulation reproduces the exact solution with high accuracy over
the range of time shown, and we again observe the growth of the
statistical errors.

\section{Conclusions}\label{CONCLU}
We have investigated two methods that yield an exact stochastic
unravelling of the non-Markovian quantum dynamics of open systems
by means of a piecewise deterministic Markov process. These
methods yield Monte Carlo simulation techniques that are generally
applicable for the investigation of the short-time behavior of the
open system's dynamics. Due to a possible exponential increase of
statistical fluctuations for large times a numerical simulation of
the long time behavior is, in general, impossible in practice.

However, a great advantage of the method is given by the fact that
it is exact and that it allows the treatment of arbitrary
correlations in the initial state. It must be emphasized that a
Monte Carlo simulation not only yields an estimate for the desired
averages, but also for the statistical errors. As long as the
latter are small the technique yields excellent predictions about
the short-time behavior and thus offers the possibility to control
and assess the performance of other approaches and approximation
schemes. In particular, the method may find important applications
in the simulation of non-Markovian decoherence phenomena which are
dominated by the short time behavior of the open system.

The formulation of the stochastic simulation method has been given
here in the interaction picture, assuming that the free dynamics
of the system and the environment are known. If this is not the
case one can use an analogous formulation of the stochastic
dynamics in the Schr\"odinger picture which includes the given
Hamiltonian operators for the system and the environment into the
deterministic drift of the stochastic differential equations
\cite{PDP-EPJD}.

In our examples we have restricted ourselves to the case of an
unpolarized (infinite temperature) initial state of the spin bath.
It should be noted that the stochastic method is also applicable
to polarized initial states. For instance, one can consider an
initial equilibrium state of finite temperature. Introducing a
spin bath Hamiltonian of the form $H_E=\omega J_3$, one then has
to multiply the probability distribution $P(j,m)$ defined in
Eq.~(\ref{PJM}) with the $m$-dependent Boltzmann factor
$\exp(-\beta\omega m)$, where $\beta$ is the inverse temperature.

There are two basic strategies for the improvement of the Monte
Carlo technique. The first one employs the freedom in the choice
of the stochastic time-evolution in order to minimize the
statistical errors \cite{LACROIX,SHAO}. This can be done by an
appropriate modification of the noise terms of the stochastic
differential equations. A further possibility is to introduce
additional terms in the deterministic part of the equations of
motion. This approach leads to a stochastic mean field dynamics
which is similar to the one used in the Monte Carlo wave function
method for interacting many-body systems \cite{CARUSO}.

The second strategy is to reduce the size of the statistical
fluctuations by using more complicated stochastic functionals
whose expectation values lead to the reduced density matrix. The
methods investigated here represent the correlated states of the
composite system through the average over random operators with a
specific given structure, namely a tensor product structure. This
ansatz does by no means exhaust all possibilities. There are many
other possible ways of constructing an exact stochastic
representation of the dynamics which seem worth being explored in
a more systematic manner.

\begin{acknowledgments}
This work was done in part at the School of Physics of the
University of KwaZulu-Natal; one of us (H.P.B.) would like to
thank the Quantum Research Group for fruitful discussions and kind
hospitality.
\end{acknowledgments}

\end{document}